\documentclass[dvipsnames, svgnames]{aastex701}

\usepackage{amsmath}
\usepackage{soul}

\soulregister\citep7
\soulregister\citet7
\soulregister\cite7

\soulregister\ref7
\soulregister\eqref7
\soulregister\pageref7

\soulregister\autoref7      
\soulregister\cref7         
\soulregister\Cref7

\sethlcolor{yellow!40}

\usepackage{graphicx} 

\shorttitle{UHECR Excesses from GRB 221009A-like Transients}
\shortauthors{Mirabal}
\begin{document}

\shorttitle{UHECR Excesses from GRB 221009A-like Transients}
\shortauthors{Mirabal}
\title{Blasts from the Past: UHECR Excesses from a Population of GRB~221009A-like Transients}

\author[orcid=0000-0002-7021-5838,sname='Mirabal']{Nestor Mirabal}
\affiliation{Center for Space Sciences and Technology, University of Maryland, Baltimore County, Baltimore, MD 21250}
\affiliation{Mail Code 661, Astroparticle Physics Laboratory, NASA Goddard Space Flight Center, Greenbelt, MD 20771, USA}
\affiliation{Center for Research and Exploration in Space Science and Technology, NASA Goddard Space Flight Center, Greenbelt, MD 20771}
\email[show]{nestor.r.mirabalbarrios@nasa.gov}

\begin{abstract}
The most prominent feature in the ultra-high-energy cosmic ray (UHECR) sky is a significant excess of events from the vicinity of Centaurus A (Cen A). 
In this paper we investigate whether the
 excess may be caused by a past GRB~221009A-like event rather than Cen A itself. If an extraordinary gamma-ray burst (GRB) went off  in one of the dwarf galaxies within the Cen A/M83 group, intergalactic magnetic deflections would naturally spread out and delay its cosmic ray signal, making it visible for $\sim 2\times 10^{5}$ years. 
 Scaling this up, we model a local population of these exceptionally bright, narrow-jet transients within the \text{Greisen--Zatsepin--Kuzmin} (GZK) horizon. For a local volumetric rate of
 $\dot n_0\simeq4\times10^{-8}\ {\rm Mpc^{-3}\,yr^{-1}}$ and a 
jet half-opening angle of $0.8^\circ$, we estimate $\simeq3-7$ UHECR excesses across the sky at any given time.
Varying the jet opening angle, GZK volume, and visibility time results in  a broader range of excess values.
It is possible to test this hypothesis by comparing the population predictions against existing and upcoming UHECR anisotropy data obtained by the Pierre Auger Observatory and the Telescope Array.
\end{abstract}



\section{Introduction} 
Because of its proximity, the nearby radio galaxy Cen A is frequently cited as the primary candidate source  for the excess of events detected in the Southern Hemisphere sky by the Pierre Auger Observatory \citep{2022ApJ...935..170A}. 
However, energetic and composition arguments suggest that the current Cen A jet activity is insufficient to accelerate particles to extreme cosmic ray energies without severe photodisintegration \citep{Liu2012}, or it requires highly specific fine tuning \citep{deOliveira2025History}. Instead, local transients such as GRBs have been proposed as the true accelerators \citep{Liu2012, Ghisellini2008UltrahighEC,Marafico_2024}. 

While a past GRB has previously been considered as a potential explanation for the 
Cen A excess \citep{PhysRevLett.105.091101}, no direct connection to a specific population of GRBs has been worked out in detail.  The discovery of the extraordinary GRB 221009A by the {\it Fermi Gamma-Ray Space Telescope}~demonstrated that exceptionally energetic GRBs can occur in the 
local Universe \citep{Burns_2023,FermiLAT_2025_BOAT}.  Dedicated modeling suggests that ultra-energetic bursts could be highly efficient accelerators capable of producing primary UHECRs and secondary gamma-ray cascades along the line of sight
\citep{AlvesBatista2022,Das2023,Mirabal2022}. While the distribution of long-duration GRBs traditionally models events such as GRB 221009A as once-in-a-10,000-year occurrences \citep{Burns_2023}, newer population models suggest that the local rate of exceptionally 
bright, narrow-jet GRB 221009A analogs
could be higher by up to two orders of magnitude $\sim 1$ per 200~years \citep{2024ApJ...975...70F}. 
Most recently, the detection of GRB 260924A provides another example of an exceptionally energetic GRB population at $z \sim 3.51$ \citep{gcn45712,gcn45713,gcn45721, gcn45740,gcn45750}.

Here we explore whether the Cen A excess could be driven by a past GRB 221009A-like transient that occurred within the Cen A/M83 group. If such extreme bursts happen intermittently throughout the local Universe, then other Earth-directed events  must also leave similar imprints across the sky. In the following sections, we calculate the expected number of localized UHECR excesses expected today and outline how to test this picture using existing and future anisotropy data.

\section{Formation of a Single UHECR Excess}
 UHECRs propagating through turbulent intergalactic magnetic fields experience spatial deflections.   For cosmic rays moving in a turbulent intergalactic magnetic field, the average deflection angle $\theta$  behaves according to \citet{Waxman:1996uz}

\begin{equation}
\theta \approx 0.54^\circ Z \left( \frac{100 \text{ EeV}}{E} \right) \left( \frac{B}{1 \text{ nG}} \right) \sqrt{\frac{D}{1 \text{ Mpc}}} \sqrt{\frac{\lambda_c}{1 \text{ Mpc}}}
\end{equation}
where $Z$ is the nuclear charge, $B$ is the magnetic field strength, $D$ is the source distance, and $\lambda_c$ is the coherence length.

Rather than predicting the angular size of an excess independently, we ask whether a physically reasonable magnetic field and nuclear composition can reproduce an angular scale comparable to the observed 
$15^\circ$ Cen A extension \citep{2022ApJ...935..170A}.
For a UHECR energy of $E=10^{20}$ eV and an iron-like composition ($Z=26$), Eq.~(1) yields
a characteristic field strength of $B\approx0.55$~nG for $D=3.8$~Mpc and $\lambda_c\sim1$~Mpc. This value is compatible with estimates of intergalactic and local-group magnetic fields \citep{2013A&ARv..21...62D}. Lighter nuclei would require a proportionally larger field to produce the same angular deflection.


Treating the observed angular scale as a characteristic propagation
deflection, the corresponding geometric delay in a turbulent magnetic field is
\begin{equation}
    \Delta t \simeq \frac{D\theta^2}{4c}.
\end{equation}
For a distance of $D\simeq3.8$ Mpc and an angular spread $\theta\simeq15^\circ$, this yields
\begin{equation}
    \Delta t \simeq 2.1\times10^{5}\ {\rm yr}.
\end{equation}

Alternatively, a cosmic ray traveling through a uniform large-scale magnetic field that bends its trajectory smoothly would roughly double this propagation time \citep{Waxman:1996uz}.
In either scenario, a GRB that occurred $\sim10^5$--$10^6$~yr ago 
could continue to feed a localized UHECR excess  today.

The previous two estimates bound a visibility window of:
\begin{equation}
\Delta t_{\rm vis}=(2-4)\times10^{5}\ {\rm yr}.
\end{equation}
This interval is a phenomenological estimate of the duration over which a transient UHECR signal contributes
to an intermediate-scale excess, rather than as a sharply defined propagation time. In a realistic propagation scenario, the observed duration depends on the distribution of the accelerated particles, their nuclear composition, the magnetic field strength, the coherence length, and the source distance. We therefore explore a broader range of the parameter space in the population calculation below. 

\begin{figure*}[ht!]
    \centering
    \plotone{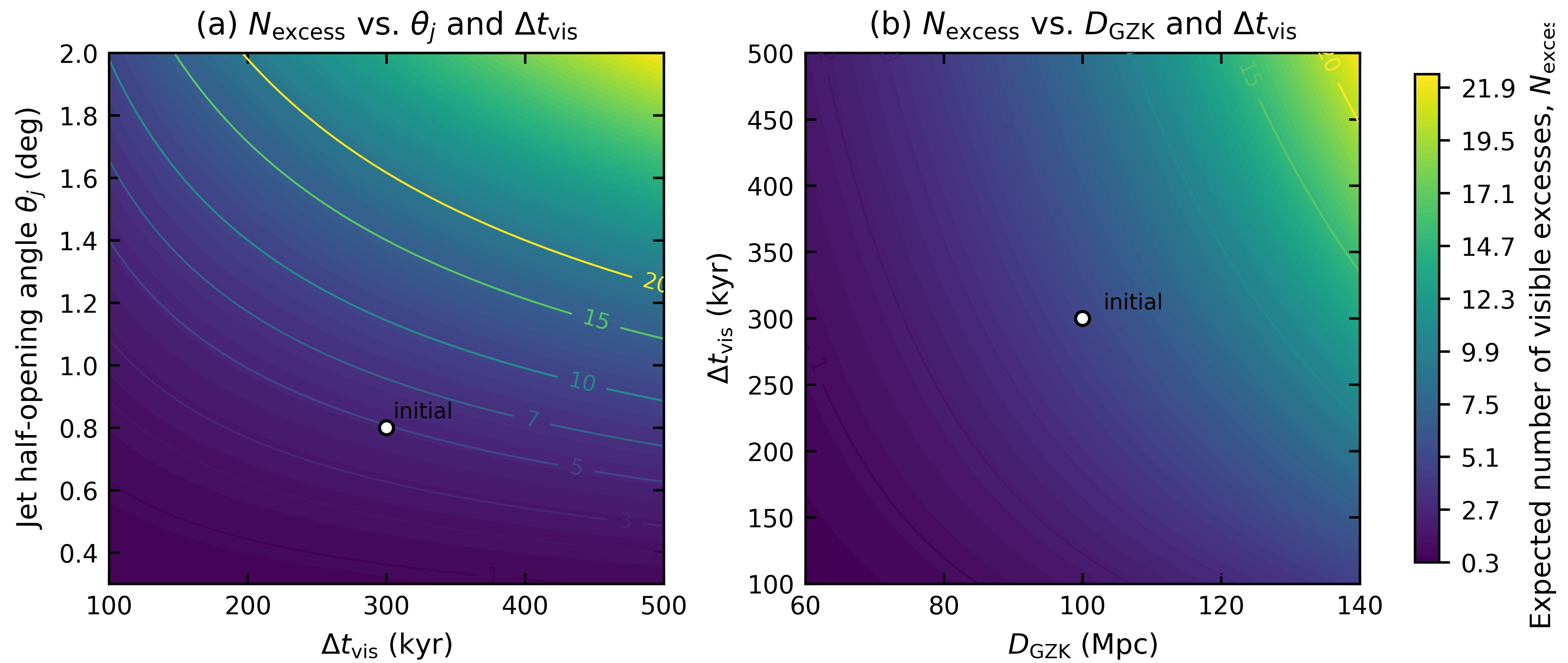}
    \caption{Number of excesses from a local population of ultra-energetic, narrow-jet GRB 221009A-like transients over a range of $\Delta t_{\rm vis}$, $\theta_{j}$ and  $D_{\rm GZK}$. The point labeled {\bf initial} marks $\Delta t_{\rm vis}$ = 300 kyr, $\theta_{j} =0.8$ and  $D_{\rm GZK} = 100$ Mpc }
    \label{fig1}
\end{figure*}

\section{Expected Number of Localized Excesses}
Following the narrow-jet formulation introduced by 
\citet{2024ApJ...975...70F}, the estimated local volumetric rate of GRB 221009A-like events is:
\begin{equation}
    \dot{n}_{0} \approx 4 \times 10^{-8} \, \text{Mpc}^{-3} \, \text{yr}^{-1}.
\end{equation}
For a characteristic UHECR horizon distance $D_{\rm GZK}$, the event
rate $R_{\rm GZK}$ within the enclosed volume is proportional to:
\begin{equation}
R_{\rm GZK}
=
\frac{4\pi}{3}D_{\rm GZK}^3\dot n_0.
\label{rgzk}
\end{equation}
Setting  $D_{\rm GZK}=100$~Mpc yields a local rate of:
\begin{equation}
R_{\rm GZK}\simeq0.17~yr^{-1}.
\label{}
\end{equation}

Only a tiny fraction of these transients will have their jet aligned toward Earth. For a population with a characteristic jet half-opening angle $\theta_j$, the beaming fraction is given by:
\begin{equation}
f_{\rm beam}=1-\cos\theta_j
\simeq\frac{\theta_j^2}{2}.
\label{beam}
\end{equation}
Assuming $\theta_j \simeq 0.8^{\circ}$
implies a beaming factor of
$f_{\rm beam}\simeq9.7\times10^{-5}$.

The expected number of UHECR excesses  across the entire sky can then be estimated by combining these terms:
\begin{equation}
   N_{\rm excess} \simeq R_{\rm GZK}\, \Delta t_{\rm vis}\, f_{\rm beam}.
\end{equation}
Adopting $R_{\rm GZK} \simeq 0.17~{\rm yr}^{-1}$ an observable cosmic ray dispersion lifetime window of $\Delta t_{\rm vis} \simeq (2-4)\times10^{5}~{\rm yr}$, we find
an expected population of $N_{\rm excess} \sim 3$--$7$ excesses  at any given time. 

The actual detectability of any individual feature will depend heavily on the event distribution, detector exposure, magnetic smearing, background fluctuations, and the angular or energy thresholds applied during the anisotropy search.
Varying the parameters across the ranges $\theta_j=0.35^\circ$--$2.0^\circ$, $D_{\rm GZK}=60$--$140$~Mpc, and $\Delta t_{\rm vis}$=$100-500$~kyr results in a broader range for $N_{\rm excess}$, 
as illustrated in Figure \ref{fig1}. 

One important caveat is that our previous calculations assume a distance-independent angular scale and visibility time. If we instead impose that $\Delta t_{\rm vis} \propto D^{2}_{\rm GZK}$ and  limit the angular size of an excess to $< 40^{\circ}$, it implies that for iron nuclei we can only resolve excesses out to a distance of $\lesssim 27$ Mpc, resulting in $N_{\rm excess} \sim 2$, at the low end of our range.

\section{Potential Connection to the Cen A Excess}
The proximity of Cen A minimizes overall propagation delays, providing an ideal local baseline to evaluate the transient scenario presented here. While the observed angular scale works reasonably well, we must also check whether the absolute energy budget of a single GRB~221009A-like transient is sufficient to power the observed excess.

\citet{2024ApJ...975...70F} evaluated the cosmic-ray energetics of the narrow-jet population directly, finding an implied local UHECR energy injection rate of $j_{\rm CR}(z\approx0)\approx4\times10^{43}\ {\rm erg\,Mpc^{-3}\,yr^{-1}}$. This can be compared with the injection rate required to sustain the observed UHECR flux against GZK losses, $j_{\rm CR,req}\approx(0.5$--$4)\times10^{44}\ {\rm erg\,Mpc^{-3}\,yr^{-1}}$ \citep{WaxmanBahcall1997,Katz2009}. The ratio of these two quantities, $\eta_{\rm CR}\equiv j_{\rm CR,req}/j_{\rm CR}\sim1$--$10$, is the baryon-loading factor the narrow-jet population would require to close the local UHECR budget  \citep{2024ApJ...975...70F}.

We can translate this population-averaged budget into a single-event energy requirement. Dividing $j_{\rm CR}$ by the adopted rate $\dot n_0\approx4\times10^{-8}\ {\rm Mpc^{-3}\,yr^{-1}}$ gives an average hadronic UHECR output per burst of
\begin{equation}
E_{\rm CR}\sim\frac{j_{\rm CR}}{\dot n_0}\approx1\times10^{51}\ {\rm erg}.
\end{equation}
This can be compared with the beaming-corrected $\gamma$-ray energy of an individual GRB~221009A-like jet, $E_\gamma=f_{\rm beam}E_{\rm iso}$, where $f_{\rm beam}=1-\cos\theta_j\simeq9.7\times10^{-5}$ for $\theta_j\simeq0.8^\circ$. Using the observed range $E_{\rm iso}\sim(3$--$10)\times10^{54}$ erg for GRB~221009A itself \citep{Burns_2023}, this gives
\begin{equation}
E_\gamma\sim(3\text{-}10)\times10^{50}\ {\rm erg}.
\end{equation}
Comparing Eqs.(10) and (11), a single Cen~A-group event would require a baryon-loading factor of $\eta_{\rm CR}=E_{\rm CR}/E_\gamma\sim1$--$3$ to supply the required population-averaged UHECR output. This is well within the $\eta \lesssim 10$ constraints inferred from  the non-detection of GRB neutrinos by IceCube \citep{IceCube2012GRB}
and the analysis of large samples of bright GRBs \citep{Zhang2025Constraints}.
This energy budget appears sufficient to sustain the cosmic ray excess observed today.

Interestingly, the environment of the Cen A/M83 group 
offers an abundance of favorable host galaxies for an extraordinary GRB 221009A-like event due to its extensive population of low-metallicity dwarf galaxies
\citep{Muller2019,Crnojevic2010,muller2025}. 
The dwarf galaxies in this group
exhibit stellar metallicities $0.03-0.08Z_\odot$, well below the canonical cutoff threshold of $Z\lesssim0.15 Z_\odot$ associated with long-duration GRB production \citep{2006AcA....56..333S}. This subsolar metallicity range closely matches the metallicity of $\approx0.12 Z_\odot$ recently measured with the \textit{James Webb Space Telescope} (\textit{JWST}) for the host galaxy of GRB 221009A \citep{Levan2024}. However, metallicity alone does not establish that such a transient
occurred  within the Cen A/M83 group. We do not claim that the Cen A excess is gamma-ray burst made, but rather view it as a potential candidate within a larger population of UHECR excesses.

\section{Discussion and Conclusions}
We propose that the Cen A excess and similar intermediate-scale anisotropies in the UHECR sky reported by the Pierre Auger Observatory and the Telescope Array may originate from past GRB~221009A analogs in the local Universe. For a $\sim 15^{\circ}$ angular scale at the distance of Cen A ($3.8$~Mpc), magnetic propagation naturally introduces a time delay of $2 \times 10^5$~yr.
Combining this timescale with the intrinsic local rate of narrow-jet GRB~221009A analogs yields an expected population of approximately $3$ to $7$ simultaneously UHECR excesses, a range that can expand under alternative parameter choices.

Searches for  anisotropies have been performed since the beginning of the Pierre Auger Observatory and the Telescope Array operations. The excess centered on Cen A is probably the most prominent one in the southern sky. But the northern sky hotspot reported by the Telescope Array has a similar intermediate-scale excess on a $\sim 20^\circ$ angular scale above 57~EeV \citep{Abbasi_2014}. Although its statistical significance and astrophysical origin remain uncertain, the existence of the latter demonstrates that comparable intermediate scale structures may be present in different regions of the UHECR sky.

While a more detailed all-sky UHECR survey remains an ongoing effort, two decades of data acquisition hints at only a handful of intermediate-scale excesses consistent with the lower range of our predictions. While still tentative, a relative low number of anisotropies without AGN counterparts would match what we would expect from highly collimated jets, where  beaming severely limits how many excesses we can see at any given time. Ongoing observations with next-generation instruments, such as AugerPrime \citep{PierreAuger:2025augerprime} and TA$\times$4\citep{TAx4_2021}, will continue to improve the precision of these anisotropy maps. Ultimately, evaluating the Cen A excess, the TA hotspot, and future directional structures will require corrections for detector exposure in both hemispheres, energy thresholds, angular smoothing, and magnetic deflections. Only then can we obtain a more definitive observational test of the transient population model.

\begin{acknowledgments}
The material is based upon work supported by NASA under award number 80GSFC24M0006.
\end{acknowledgments}

%
\facilities{Pierre Auger Observatory, Telescope Array, Fermi Gamma-Ray Space Telescope}

\software{Astropy\citep{astropy:2013}}


\bibliography{ref}{}

\begin{thebibliography}{}
\expandafter\ifx\csname natexlab\endcsname\relax\def\natexlab#1{#1}\fi
\providecommand{\url}[1]{\href{#1}{#1}}
\providecommand{\dodoi}[1]{doi:~\href{http://doi.org/#1}{\nolinkurl{#1}}}
\providecommand{\doeprint}[1]{\href{http://ascl.net/#1}{\nolinkurl{http://ascl.net/#1}}}
\providecommand{\doarXiv}[1]{\href{https://arxiv.org/abs/#1}{\nolinkurl{https://arxiv.org/abs/#1}}}

\bibitem[{R.~U. Abbasi {et~al.}(2014)Abbasi {et~al.}}]{Abbasi_2014}
Abbasi, R.~U., {et~al.} 2014, \bibinfo{title}{Indications of Intermediate-Scale
  Anisotropy of Ultrahigh-Energy Cosmic Rays in the Northern Sky,} The
  Astrophysical Journal, 790, L21, \dodoi{10.1088/2041-8205/790/2/L21}

\bibitem[{P. {Abreu} {et~al.}(2022){Abreu}, {Aglietta}, {Albury}, {Allekotte},
  {Almeida Cheminant}, {Almela}, {Alvarez-Mu{\~n}iz}, {Alves Batista},
  {Ammerman Yebra}, {Anastasi}, {Anchordoqui}, {Andrada}, {Andringa}, {Aramo},
  {Ara{\'u}jo Ferreira}, {Arnone}, {Arteaga Vel{\'a}zquez}, {Asorey}, {Assis},
  {Avila}, {Avocone}, {Badescu}, {Bakalova}, {Balaceanu}, {Barbato}, {Bellido},
  {Berat}, {Bertaina}, {Bhatta}, {Biermann}, {Binet}, {Bismark}, {Bister},
  {Biteau}, {Blazek}, {Bleve}, {Bl{\"u}mer}, {Boh{\'a}{\v{c}}ov{\'a}},
  {Boncioli}, {Bonifazi}, {Bonneau Arbeletche}, {Borodai}, {Botti}, {Brack},
  {Bretz}, {Brichetto Orchera}, {Briechle}, {Buchholz}, {Bueno}, {Buitink},
  {Buscemi}, {B{\"u}sken}, {Caballero-Mora}, {Caccianiga}, {Canfora},
  {Caracas}, {Caruso}, {Castellina}, {Catalani}, {Cataldi}, {Cazon}, {Cerda},
  {Chinellato}, {Chudoba}, {Chytka}, {Clay}, {Cobos Cerutti}, {Colalillo},
  {Coleman}, {Coluccia}, {Concei{\c{c}}{\~a}o}, {Condorelli}, {Consolati},
  {Contreras}, {Convenga}, {Correia dos Santos}, {Covault}, {Dasso},
  {Daumiller}, {Dawson}, {Day}, {de Almeida}, {de Jes{\'u}s}, {de Jong}, {de
  Mello Neto}, {De Mitri}, {de Oliveira}, {de Oliveira Franco}, {de Palma}, {de
  Souza}, {De Vito}, {Del Popolo}, {del R{\'\i}o}, {Deligny}, {Deval}, {di
  Matteo}, {Dobre}, {Dobrigkeit}, {D'Olivo}, {Domingues Mendes}, {dos Anjos},
  {Dova}, {Ebr}, {Engel}, {Epicoco}, {Erdmann}, {Escobar}, {Etchegoyen},
  {Falcke}, {Farmer}, {Farrar}, {Fauth}, {Fazzini}, {Feldbusch}, {Fenu},
  {Fick}, {Figueira}, {Filip{\v{c}}i{\v{c}}}, {Fitoussi}, {Fodran}, {Fujii},
  {Fuster}, {Galea}, {Galelli}, {Garc{\'\i}a}, {Gemmeke}, {Gesualdi},
  {Gherghel-Lascu}, {Ghia}, {Giaccari}, {Giammarchi}, {Glombitza}, {Gobbi},
  {Gollan}, {Golup}, {G{\'o}mez Berisso}, {G{\'o}mez Vitale}, {Gongora},
  {Gonz{\'a}lez}, {Gonz{\'a}lez}, {Goos}, {G{\'o}ra}, {Gorgi}, {Gottowik},
  {Grubb}, {Guarino}, {Guedes}, {Guido}, {Hahn}, {Hamal}, {Hampel}, {Hansen},
  {Harari}, {Harvey}, {Haungs}, {Hebbeker}, {Heck}, {Hill}, {Hojvat},
  {H{\"o}randel}, {Horvath}, {Hrabovsk{\'y}}, {Huege}, {Insolia}, {Isar},
  {Janecek}, {Johnsen}, {Jurysek}, {K{\"a}{\"a}p{\"a}}, {Kampert}, {Keilhauer},
  {Khakurdikar}, {Kizakke Covilakam}, {Klages}, {Kleifges}, {Kleinfeller},
  {Knapp}, {Kunka}, {Lago}, {Langner}, {Leigui de Oliveira}, {Lenok},
  {Letessier-Selvon}, {Lhenry-Yvon}, {Lo Presti}, {Lopes}, {L{\'o}pez}, {Lu},
  {Luce}, {Lundquist}, {Machado Payeras}, {Mancarella}, {Mandat}, {Manning},
  {Manshanden}, {Mantsch}, {Marafico}, {Mariani}, {Mariazzi}, \&
  {Mari{\textcommabelow s}}}]{2022ApJ...935..170A}
{Abreu}, P., {Aglietta}, M., {Albury}, J.~M., {et~al.} 2022,
  \bibinfo{title}{{Arrival Directions of Cosmic Rays above 32 EeV from Phase
  One of the Pierre Auger Observatory},} \apj, 935, 170,
  \dodoi{10.3847/1538-4357/ac7d4e}

\bibitem[{M. {Ajello} {et~al.}(2025){Ajello}, {Baldini}, {Barbiellini},
  {et~al.}}]{FermiLAT_2025_BOAT}
{Ajello}, M., {Baldini}, L., {Barbiellini}, G., {et~al.} 2025,
  \bibinfo{title}{GRB 221009A: The B.O.A.T. Burst that Shines in Gamma Rays,}
  The Astrophysical Journal Supplement Series, 277, 24,
  \dodoi{10.3847/1538-4365/ada272}

\bibitem[{R. Alves~Batista(2022)Alves~Batista}]{AlvesBatista2022}
Alves~Batista, R. 2022, \bibinfo{title}{{GRB 221009A: a potential source of
  ultra-high-energy cosmic rays},} arXiv e-prints, arXiv:2210.12855,
  \dodoi{10.48550/arXiv.2210.12855}

\bibitem[{ {Astropy Collaboration} {et~al.}(2013){Astropy Collaboration},
  {Robitaille}, {Tollerud}, {Greenfield}, {Droettboom}, {Bray}, {Aldcroft},
  {Davis}, {Ginsburg}, {Price-Whelan}, {Kerzendorf}, {Conley}, {Crighton},
  {Barbary}, {Muna}, {Ferguson}, {Grollier}, {Parikh}, {Nair}, {Unther},
  {Deil}, {Woillez}, {Conseil}, {Foley}, {Kuusk}, {Robitaille}, {Feiler},
  {Joseph}, {他說}, {Markov}, {Greenfield}, {peyther}, {Sepulveda},
  {Golkhou}, {Pease}, {Jannuzi}, {Schilling}, {Rest}, {Sick}, {Rojas},
  {Rosenfield}, {Roman M.}, {Craig}, {Zhu}, {Bach}, {Duzhev}, {Maraganore},
  {Levine}, {Churazov}, \& {Denk}}]{astropy:2013}
{Astropy Collaboration}, {Robitaille}, T.~P., {Tollerud}, E.~J., {et~al.} 2013,
  \bibinfo{title}{{Astropy: A community Python package for astronomy},} A\&A,
  558, A33, \dodoi{10.1051/0004-6361/201322068}

\bibitem[{E. Burns {et~al.}(2023)Burns, Svinkin, Fenimore,
  {et~al.}}]{Burns_2023}
Burns, E., Svinkin, D., Fenimore, E., {et~al.} 2023, \bibinfo{title}{GRB
  221009A: The B.O.A.T.,} The Astrophysical Journal Letters, 946, L31,
  \dodoi{10.3847/2041-8213/acc39c}

\bibitem[{A. Calvez {et~al.}(2010)Calvez, Kusenko, \&
  Nagataki}]{PhysRevLett.105.091101}
Calvez, A., Kusenko, A., \& Nagataki, S. 2010, \bibinfo{title}{Role of Galactic
  Sources and Magnetic Fields in Forming the Observed Energy-Dependent
  Composition of Ultrahigh-Energy Cosmic Rays,} Phys. Rev. Lett., 105, 091101,
  \dodoi{10.1103/PhysRevLett.105.091101}

\bibitem[{A.~J. {Castro-Tirado} {et~al.}(2026){Castro-Tirado}, {Perez-Garcia},
  {Sota}, {Fernandez-Garcia}, {Caballero-Garcia}, {Guziy}, {Gritsevich}, {Hu},
  {Fernandez-Martin}, \& {Gongora}}]{gcn45740}
{Castro-Tirado}, A.~J., {Perez-Garcia}, I., {Sota}, A., {et~al.} 2026,
  \bibinfo{title}{{GRB 260924A: 2.2m CAHA spectroscopic redshift z = 3.51},}
  GCN Circular, 45740, 1

\bibitem[{D. {Crnojevi{\'c}} {et~al.}(2010){Crnojevi{\'c}}, {Grebel}, \&
  {Koch}}]{Crnojevic2010}
{Crnojevi{\'c}}, D., {Grebel}, E.~K., \& {Koch}, A. 2010, \bibinfo{title}{{A
  close look at the Centaurus A group of galaxies. I. Metallicity distribution
  functions and population gradients in early-type dwarfs},} A\&A, 516, A85,
  \dodoi{10.1051/0004-6361/200913210}

\bibitem[{S. {Das} \& S. {Razzaque}(2023){Das} \& {Razzaque}}]{Das2023}
{Das}, S., \& {Razzaque}, S. 2023, \bibinfo{title}{{Ultrahigh-energy cosmic-ray
  signature in GRB 221009A},} A\&A, 670, L12,
  \dodoi{10.1051/0004-6361/202245377}

\bibitem[{C.~a. de~Oliveira \& V. de~Souza(2026)de~Oliveira \&
  de~Souza}]{deOliveira2025History}
de~Oliveira, C.~a., \& de~Souza, V. 2026, \bibinfo{title}{{History of UHECR
  production in Centaurus A},} Monthly Notices of the Royal Astronomical
  Society, 545, staf2160, \dodoi{10.48550/arXiv.2507.06894}

\bibitem[{V. {D'Elia} {et~al.}(2026){D'Elia}, {DeLaunay}, {Gupta}, {Lien},
  {Page}, {Palmer}, {Parsotan}, \& {Sbarufatti}}]{gcn45713}
{D'Elia}, V., {DeLaunay}, J.~J., {Gupta}, R., {et~al.} 2026,
  \bibinfo{title}{{GRB 260924A: Swift detection of a burst},} GCN Circular,
  45713, 1

\bibitem[{R. {Durrer} \& A. {Neronov}(2013){Durrer} \&
  {Neronov}}]{2013A&ARv..21...62D}
{Durrer}, R., \& {Neronov}, A. 2013, \bibinfo{title}{{Cosmological magnetic
  fields},} The Astronomy and Astrophysics Review, 21, 62,
  \dodoi{10.1007/s00159-013-0062-7}

\bibitem[{ {Fermi GBM Team}(2026){Fermi GBM Team}}]{gcn45712}
{Fermi GBM Team}. 2026, \bibinfo{title}{{GRB 260924A: Fermi GBM Final Real-time
  Localization},} GCN Circular, 45712, 1

\bibitem[{J.~D. {Finke} \& S. {Razzaque}(2024){Finke} \&
  {Razzaque}}]{2024ApJ...975...70F}
{Finke}, J.~D., \& {Razzaque}, S. 2024, \bibinfo{title}{{Is Gamma-Ray Burst
  221009A Really a Once-in-10,000 yr Event?},} \apj, 975, 70,
  \dodoi{10.3847/1538-4357/ad78dd}

\bibitem[{G. Ghisellini {et~al.}(2008)Ghisellini, Ghirlanda, Tavecchio,
  Fraternali, Ghisellini, {et~al.}}]{Ghisellini2008UltrahighEC}
Ghisellini, G., Ghirlanda, G., Tavecchio, F., {et~al.} 2008,
  \bibinfo{title}{Ultra-high energy cosmic rays, spiral galaxies and
  magnetars,} Monthly Notices of the Royal Astronomical Society: Letters, 390,
  L88, \dodoi{10.1111/j.1745-3933.2008.00547.x}

\bibitem[{R. {Gupta} {et~al.}(2026){Gupta}, {Veres}, {Chand}, {Burns},
  {McDermott}, \& {Scotton}}]{gcn45750}
{Gupta}, R., {Veres}, P., {Chand}, V., {et~al.} 2026, \bibinfo{title}{{GRB
  260924A: Extreme Isotropic-equivalent Gamma-ray Energy and encouragement for
  multiwavelength follow-up observations},} GCN Circular, 45750, 1

\bibitem[{A. {Holzmann Airasca} {et~al.}(2026){Holzmann Airasca}, {Depalo},
  {Cutini}, {Khalil}, \& {Gupta}}]{gcn45721}
{Holzmann Airasca}, A., {Depalo}, D., {Cutini}, S., {Khalil}, T., \& {Gupta},
  R. 2026, \bibinfo{title}{{GRB 260924A: Fermi-LAT detection of a bright
  burst},} GCN Circular, 45721, 1

\bibitem[{ {IceCube Collaboration} {et~al.}(2012){IceCube Collaboration},
  Abbasi, {et~al.}}]{IceCube2012GRB}
{IceCube Collaboration}, Abbasi, R., {et~al.} 2012, \bibinfo{title}{An absence
  of neutrinos associated with cosmic-ray acceleration in gamma-ray bursts,}
  Nature, 484, 351, \dodoi{10.1038/nature11042}

\bibitem[{B. {Katz} {et~al.}(2009){Katz}, {Budnik}, \& {Waxman}}]{Katz2009}
{Katz}, B., {Budnik}, R., \& {Waxman}, E. 2009, \bibinfo{title}{{The energy
  production rate and generation spectrum of UHECRs},} Journal of Cosmology and
  Astroparticle Physics, 2009, 020, \dodoi{10.1088/1475-7516/2009/03/020}

\bibitem[{E. Kido {et~al.}(2021)Kido {et~al.}}]{TAx4_2021}
Kido, E., {et~al.} 2021, \bibinfo{title}{Surface detectors of the TAx4
  experiment,} Nuclear Instruments and Methods in Physics Research Section A:
  Accelerators, Spectrometers, Detectors and Associated Equipment, 1021,
  165910, \dodoi{10.1016/j.nima.2021.165910}

\bibitem[{A.~J. {Levan} {et~al.}(2024){Levan}, {Lamb}, {Orchard},
  {et~al.}}]{Levan2024}
{Levan}, A.~J., {Lamb}, G.~P., {Orchard}, M.~J., {et~al.} 2024,
  \bibinfo{title}{{JWST detection of a supernova associated with GRB 221009A
  without an r-process signature},} Nature Astronomy, 8, 712,
  \dodoi{10.1038/s41550-024-02237-4}

\bibitem[{R.-Y. Liu {et~al.}(2012)Liu, Wang, Wang, \& Taylor}]{Liu2012}
Liu, R.-Y., Wang, X.-Y., Wang, W., \& Taylor, A.~M. 2012, \bibinfo{title}{On
  the excess of ultra-high energy cosmic rays in the direction of Centaurus A,}
  The Astrophysical Journal, 744, 73, \dodoi{10.1088/0004-637X/744/1/73}

\bibitem[{S. Marafico {et~al.}(2024)Marafico, Biteau, Condorelli, Deligny, \&
  Bregeon}]{Marafico_2024}
Marafico, S., Biteau, J., Condorelli, A., Deligny, O., \& Bregeon, J. 2024,
  \bibinfo{title}{Closing the Net on Transient Sources of Ultra-high-energy
  Cosmic Rays,} The Astrophysical Journal, 972, 148,
  \dodoi{10.3847/1538-4357/ad5a11}

\bibitem[{N. {Mirabal}(2023){Mirabal}}]{Mirabal2022}
{Mirabal}, N. 2023, \bibinfo{title}{{Secondary GeV--TeV emission from
  ultra-high-energy cosmic rays in GRB 221009A},} Mon. Not. R. Astron. Soc.
  Lett., 519, L85, \dodoi{10.1093/mnrasl/slac149}

\bibitem[{O. {M{\"u}ller} {et~al.}(2019){M{\"u}ller}, {Rejkuba}, {Jerjen},
  {et~al.}}]{Muller2019}
{M{\"u}ller}, O., {Rejkuba}, M., {Jerjen}, H., {et~al.} 2019,
  \bibinfo{title}{{The dwarf galaxy satellite system of Centaurus A},} A\&A,
  629, A18, \dodoi{10.1051/0004-6361/201935411}

\bibitem[{O. {M{\"u}ller} {et~al.}(2025){M{\"u}ller}, {Rejkuba}, {Fahrion},
  {Pawlowski}, {Famaey}, {Libeskind}, {Heesters}, {Federico}, {Hilker},
  {Taibi}, \& {Pearson}}]{muller2025}
{M{\"u}ller}, O., {Rejkuba}, M., {Fahrion}, K., {et~al.} 2025,
  \bibinfo{title}{{MUSE observations of dwarf galaxies and a stellar stream in
  the M 83 group},} \textsl{Astronomy \& Astrophysics}, 699, A207,
  \dodoi{10.1051/0004-6361/202552923}

\bibitem[{ {Pierre Auger Collaboration}(2025){Pierre Auger
  Collaboration}}]{PierreAuger:2025augerprime}
{Pierre Auger Collaboration}. 2025, \bibinfo{title}{{AugerPrime: Status and
  first results},} Proceedings of Science (PoS), 39th International Cosmic Ray
  Conference (ICRC2025), 385.
\newblock \doarXiv{2508.08056}

\bibitem[{K.~Z. {Stanek} {et~al.}(2006){Stanek}, {Gnedin}, {Beacom}, {Gould},
  {Johnson}, {Kollmeier}, {Modjaz}, {Pinsonneault}, {Pogge}, \&
  {Weinberg}}]{2006AcA....56..333S}
{Stanek}, K.~Z., {Gnedin}, O.~Y., {Beacom}, J.~F., {et~al.} 2006,
  \bibinfo{title}{{Protecting Life in the Milky Way: Metals Keep the GRBs
  Away},} \actaa, 56, 333, \dodoi{10.48550/arXiv.astro-ph/0604113}

\bibitem[{E. {Waxman} \& J. {Bahcall}(1997){Waxman} \&
  {Bahcall}}]{WaxmanBahcall1997}
{Waxman}, E., \& {Bahcall}, J. 1997, \bibinfo{title}{{High Energy Neutrinos
  from Cosmological Gamma-Ray Burst Fireballs},} Physical Review Letters, 78,
  2292, \dodoi{10.1103/PhysRevLett.78.2292}

\bibitem[{E. Waxman \& J. Miralda-Escud{\'e}(1996)Waxman \&
  Miralda-Escud{\'e}}]{Waxman:1996uz}
Waxman, E., \& Miralda-Escud{\'e}, J. 1996, \bibinfo{title}{{Images of Bursting
  Sources of High-Energy Cosmic Rays. I. Effects of Magnetic Fields},}
  Astrophys. J. Lett., 472, L89, \dodoi{10.1086/310367}

\bibitem[{B.~B. Zhang {et~al.}(2025)Zhang {et~al.}}]{Zhang2025Constraints}
Zhang, B.~B., {et~al.} 2025, \bibinfo{title}{Constraints on Cosmic-Ray
  Acceleration in Bright Gamma-Ray Bursts,} The Astrophysical Journal, 981, 45,
  \dodoi{10.3847/1538-4357/ada941}

\end{thebibliography}
\bibliographystyle{aasjournalv7}



\end{document}